\documentclass[conference]{IEEEtran}

\usepackage{ifpdf}
\ifpdf

\ifCLASSOPTIONcompsoc
  \usepackage[nocompress]{cite}
\else
  \usepackage{cite}
\fi
\ifCLASSINFOpdf \else \fi

\usepackage{amsmath}
\usepackage{algorithmic}
\usepackage{array}
\usepackage{bm}
\usepackage[pdftex]{graphicx}
\usepackage{subfigure}
\usepackage{multirow}
\usepackage{booktabs}
\usepackage{indentfirst}
\usepackage{amsfonts,amssymb}
\usepackage{lineno}
\usepackage{color}
\usepackage{algorithm}
\usepackage{algorithmic}
\usepackage{threeparttable}

\renewcommand\thesection{\Roman{section}}

\begin{document}

\title{Soft MIMO Detection Using Marginal Posterior Probability Statistics}

\author{
	\centerline{Jiankun Zhang, Hao Wang, Jing Qian, and Zhenxing Gao}\\
    \IEEEauthorblockA{
        Huawei Technologies Co., Ltd, Beijing 100095, China\\
		Email: \{zhangjiankun4, hunter.wanghao, qianjing3, gaozhenxing\}@huawei.com}
}

\maketitle

\begin{abstract}
Soft demodulation of received symbols into bit log-likelihood ratios (LLRs) is at the very heart of multiple-input-multiple-output (MIMO) detection. However, the optimal maximum a posteriori (MAP) detector is complicated and infeasible to be used in a practical system. In this paper, we propose a soft MIMO detection algorithm based on marginal posterior probability statistics (MPPS). With the help of optimal transport theory and order statistics theory, we transform the posteriori probability distribution of each layer into a Gaussian distribution. Then the full sampling paths can be implicitly restored from the first- and second-order moment statistics of the transformed distribution. A lightweight network is designed to learn to recovery the log-MAP LLRs from the moment statistics with low complexity. Simulation results show that the proposed algorithm can improve the performance significantly with reduced samples under fading and correlated channels.
\end{abstract}

\IEEEpeerreviewmaketitle

\section{Introduction}

Multiple-input-multiple-output (MIMO) systems have great potential to increase spectral efficiency by transmitting independent data streams on multiple antennas, which has become an important technology for 5G and beyond. As the computational complexity increases exponentially with the number of antennas, the optimal soft-decision detector, the maximum a posteriori (MAP) detector, will consume enormous computing power and require tremendous computational resources. MAP detector is infeasible to be used even for practical small antenna systems such as user equipment.

Different approaches have been proposed to address the MIMO detection problem. One type of the most investigated algorithms is the quasi-maximum likelihood (ML) detector based on tree-search. It interprets the detection as the search for the best path in a tree, and aims to select an efficient subset of the exhaustive search as a candidate list, based on which both the hard output and the log-likelihood ratio (LLR) can be calculated. A low complexity quasi-ML detection was investigated using sphere decoding (SD) for soft detection [1]. More methods are designed to pursue minimization of the number of path that is needed to obtain the ML solution. Among them, the fixed-complexity sphere decoder (FSD) [2], imbalanced FSD (IFSD) [3], parallel SD [4], and random sampling (RS) [5] are proposed to optimize the searching space for ML solution at low cost. Recently, machine learning approaches such as DetNet [6] and LISA [7] are proposed to unfold gradient descent like solution for the ML optimization.
However, ML detector is initially designed for hard decision systems. For soft MIMO detection, in addition to the ML solution, a set of searches for potential counter-hypotheses are also required for tree-search algorithms to get the soft outputs. The counter-hypotheses of certain bits may be missing due to tree pruning, thus the LLR values for certain bits cannot be calculated correctly which leads to further performance degradation [8].

Aiming to the above shortcomings, there are some methods to optimize the soft quasi-ML detection. [9,10] proposed the smart ordering and candidate adding (SOCA) method which approximates the LLR by flipping the desired bit of the ML hypothesis. [11] proposed an improved LLR approximation algorithm by applying the metric threshold of the missing counter-hypothesis using partial metrics. [12] proposed a reliability-aware transformation technique, which exploits the statistical relationship between the raw LLR's amplitude and its reliability to adjust the raw LLR's amplitude. [13] proposed a trainable iterative detector based on the iterative soft thresholding algorithm (ISTA). However, all these methods either need to expand the searching space, or suffer from the block-error-rate (BLER) performance degradation.

Another type of soft MIMO detection is to relax the discrete random variables to be continuous and assume a Gaussian-distributed prior probability, which can be exempt from the need of search for counter-hypotheses. One remarkable example is the linear minimum mean square error (LMMSE) equalization [14]. But linear equalization with subsequent detection leads to a strong performance degradation. Besides, message passing  based algorithms, such as approximate message passing (AMP) [15] and expectation propagation (EP) [16], have been proposed for MIMO detection. Machine learning approaches such as OAMP-Net [17] are designed by unfolding the orthogonal AMP (OAMP) algorithm. These methods are derived by approximating the posterior distribution with factorized Gaussian distributions and can achieve Bayes-optimal performance in large-scale MIMO systems. However, for practical small-size (e.g., 4x4) MIMO systems, the performance of these iterative detectors is still far from Bayes-optimal solution and has serious deterioration under correlated MIMO channels [17,18].

In this paper, we propose a novel soft MIMO detection approach based on observed marginal posterior probability statistics (MPPS), which combines the advantages of the above two types of the soft MIMO detectors. Unlike the aforementioned algorithms, we fit the MAP distribution of each layer of the received signal by a small number of sampling paths. Since the discrete marginal probability distribution of each layer is not Gaussian distributed, we transform the path metrics of an arbitrary distribution into an unimodal distribution under the guidance of optimal transport theory [19] and order statistics theory [20], which aims to achieve the smallest Kullback-Leibler (KL) divergence between the transformed distribution and a Gaussian distribution. Using the statistical characteristics of sampling paths, only the moments of each layer are stored and the full sampling path is implicitly restored from the moment features by a machine learning approach, and the log-MAP LLR is therefore calculated without the issue of missing counter-hypotheses. Simulation results show that MPPS can achieve significant gain in performance compared with traditional quasi-ML detector with low complexity.

\section{System Model}


Let us consider a MIMO downlink communication system where the received signal can be modeled as
\begin{equation}
\displaystyle{\mathbf{y}=\mathbf{H}\mathbf{s}+\mathbf{n},}
\end{equation}
where $\mathbf{H} \in \mathbb{C}^{N_r \times N_t}$ is the channel gain matrix, $\mathbf{s}(n) = [s_1(n), , s_2(n), · · · , s_{N_t}(n)]^T$ is the transmit symbol containing original bits, where $N_t$ and $N_r$ are the number of transmitting and receiving antennas, respectively. The complex additive white Gaussian noise obeys $\mathbf{n}\sim \mathcal{C}\mathcal{N}\left( \mathbf{0},~{{\sigma }^{2}}\mathbf{I} \right)$. At the soft demodulator, the LLR of the $i$-th bit in the $j$-th layer is expressed as
\begin{equation}
\displaystyle{{\lambda_{b_{j}^{i}}}=\log \frac{p(b_{j}^{i}=1|\mathbf{y})}{p(b_{j}^{i}=0|\mathbf{y})}=\log \frac{\mathop{\sum }_{\mathbf{s}\in {{\mathcal{S}}_j^{\left( 1 \right)}}}p\left(\mathbf{s}|\mathbf{y} \right)}{\mathop{\sum }_{\mathbf{s}\in {{\mathcal{S}}_j^{\left( 0 \right)}}}p\left( \mathbf{s}|\mathbf{y} \right)},}
\end{equation}
where $b_{j}^{i}$ is the $i$-th bit in the $j$-th dimension, $\mathcal{S}_j^{(1)}$ is the assemble of $\mathbf{s}$ satisfying $b_j^i=1$, $\mathcal{S}_j^{(0)}$ is the assemble of $\mathbf{s}$ satisfying $b_j^i=0$, $p(\mathbf{s}|\mathbf{y})$ is a posteriori probability of the transmitted symbol. According to Bayesian theory, the posterior probability of received symbols can be transformed into the following expression of likelihood probability
\begin{equation}
\displaystyle{p\left( \mathbf{s}|\mathbf{y} \right)=\frac{p\left( \mathbf{y}|\mathbf{s} \right)}{\sum\limits_{\mathbf{s}\in \mathcal{S}}p\left( \mathbf{y}|\mathbf{s} \right)}=\frac{\exp \left( -{{||\mathbf{y} - \mathbf{H}\mathbf{s}||}^{2}} \right)}{\sum\limits_{\mathbf{s}\in \mathcal{S}}\exp \left( -{{||\mathbf{y} - \mathbf{H}\mathbf{s}||}^{2}} \right)},}
\end{equation}
where $p(\mathbf{y}|\mathbf{s})$ is the likelihood probability. By substituting (3) into (2) and using max-log approximation [3], LLR can be written as
\begin{equation}
\displaystyle{{{\lambda }_{b_{j}^{i}}}\approx \frac{1}{2{{\sigma }^{2}}}\left( \underset{\mathbf{s}\in {{\mathcal{S}}^{\left( 0 \right)}}}{\mathop{\min }}\,{{||\mathbf{y} - \mathbf{H}\mathbf{s}||}^{2}}-\underset{\mathbf{s}\in {{\mathcal{S}}^{\left( 1 \right)}}}{\mathop{\min }}\,{{||\mathbf{y} - \mathbf{H}\mathbf{s}||}^{2}} \right).}
\end{equation}
If we call the vector which has the minimum metric as ML hypothesis and the vectors which have at least one different bit from the ML hypothesis as counter-hypotheses, then the LLR value is the difference between the metric of the ML hypothesis and that of the counter-hypothesis. However, in the soft-output tree searching algorithm, if the searching space is not large enough, we may find sometimes all the candidates have the same bit at a certain position, which is called missing counter-hypotheses problem. A soft MIMO detection algorithm will be proposed to solve this issue.

\section{Theoretical Principle}

Traditional tree search based ML detection schemes have two challenges. First, the number of sampling paths, storage complexity, and operation complexity all increase with the transmission symbol dimension $N_t$ and the transmission symbol constellation point size $M_c$. For higher-order modulation modes, the complexity will increase significantly. Second, ML detection is a multi-mode discrete optimization problem. The algorithm based on gradient descent is easy to fall into the local optimal solution. The impact is more obvious for greater channel correlation and higher modulation order.

To solve the problems of increasing searching space and missing counter-hypotheses, we propose a new idea to estimate the statistical characteristics of $p(\mathbf{s}|\mathbf{y})$  through the sampled paths. According to (3), $p(\mathbf{s}|\mathbf{y})$  can be represented by the path metric of the sampled paths. With a number of sampled paths, $p(\mathbf{s}|\mathbf{y})$ can be fitted.

\subsection{Fitting the marginal posterior probability}

Here, we can rewrite (3) as
\begin{small}
\begin{equation}
\displaystyle{\log p(\mathbf{s} \mid \mathbf{y})=-\|\mathbf{y}-\mathbf{H} \mathbf{s}\|^{2}-\log \sum_{s \in \mathcal{S}} \exp \left(-\|\mathbf{y}-\mathbf{H} \mathbf{s}\|^{2}\right).}
\end{equation}
\end{small}
The marginal posteriori probability of a specific symbol $s_j$, which is the $j$-th component of $\mathbf{s}$, can be written as
\begin{equation}
\displaystyle{p\left(s_{j}=X_{i} \mid \mathbf{y}\right)=\sum_{\mathbf{s} \in\left\{s_{j}=X_{i}\right\}} p(\mathbf{s} \mid \mathbf{y}),}
\end{equation}
where $X_i$ denotes the $i$-th constellation point, with $i=0,1,…,2^{M_c}-1$. Substituting (5) into (6) and using  max-log approximation, we can derive
\begin{equation}
\begin{aligned}
\log p\left(s_{j}=X_{i} \mid \mathbf{y}\right) & \approx-\min _{\mathbf{s} \in\left\{s_{i}=X_{i}\right\}}\|\mathbf{y}-\mathbf{H} \mathbf{s}\|^{2} \\
&+\min _{\mathbf{s} \in \mathcal{S}}\|\mathbf{y}-\mathbf{H} \mathbf{s}\|^{2} .
\end{aligned}
\end{equation}

The physical meaning of max-log approximation in (7) is to approximate the projection of Gaussian distribution as a section along the distributed ridge. When the high-dimensional distribution conforms to the joint multi-dimensional Gaussian distribution (corresponding to low correlation between layers), the accuracy of max-log-MAP approximation is high because the distribution of ridges is consistent with the projection distribution. When the received signal-to-noise ratio (SNR) is high, the accuracy of max-log-MAP approximation is also high because the probability value is concentrated near the maximum point.

In the following, we show the derivations of the posteriori moments (mean and variance) from the sampled path metrics. if we assume $p(s_{j} \mid \mathbf{y})$ follows Gaussian distribution with a mean $\mu_j$ and a variance $\sigma_j^2$, i.e. $p(s_{j} \mid \mathbf{y})\sim\mathcal{N}(\mu_j,\sigma_j^2)$, we can derive
\begin{equation}
\displaystyle{\log p\left(s_{j}=X_{i} \mid \mathbf{y}\right)=-\frac{\left|X_{i}-\mu_{j}\right|^{2}}{2 \sigma_{j}^{2}}-\frac{1}{2} \log \left(2 \pi \sigma_{j}^{2}\right).}
\end{equation}
From (7) and (8), we can get
\begin{equation}
\begin{aligned}
\min _{s \in\left\{s_{j}=X_{i}\right\}}\|\mathbf{y}-\mathbf{H} \mathbf{s}\|^{2} &=\frac{\left|X_{i}-\mu_{j}\right|^{2}}{2 \sigma_{j}^{2}} \\
&+\frac{1}{2} \log \left(2 \pi \sigma_{j}^{2}\right)+\min _{s \in \mathcal{S}}\|\mathbf{y}-\mathbf{H s}\|^{2}
\end{aligned}
\end{equation}
Considering that the real and imaginary part of $s_j$ is orthogonal within each layer, in order to simplify the derivations without loss of generality, we consider the one-dimensional case, that is, $s_j$ is a real number representing either the real part or imaginary part of constellation points. Here we assume the standard quadrature amplitude modulation (QAM) and suppose $X_i=2(i-M)+1,i=0,1,\cdots,2M-1$. If we denote $D_{i,j}=\min _{s \in\left\{s_{j}=X_{i}\right\}}\|\mathbf{y}-\mathbf{H} \mathbf{s}\|^{2}$, then $D_{i,j}-D_{i+1,j}$ is given by
\begin{equation}
\displaystyle{D_{i, j}-D_{i+1, j}=\frac{1}{2 \sigma_{j}^{2}}\left(X_{i+1}^{2}-X_{i}^{2}\right)+\frac{1}{\sigma_{j}^{2}}\left(X_{i}-X_{i+1}\right) \mu_{j},}
\end{equation}
which can be further simplified as
\begin{equation}
\displaystyle{D_{i, j}-D_{i+1, j}=\frac{2}{\sigma_{j}^{2}} X_{i}+\frac{2}{\sigma_{j}^{2}}\left(1-\mu_{j}\right).}
\end{equation}
Here we have used the relationship $X_{i+1}^2-X_i^2=4(X_i+1)$.

In the above equation, $\{X_i,D_{i,j}\}$ are known parameters that can be measured, and $\mu_j,\sigma_j^2$ are unknown parameters to be solved. Since $D_{i,j}-D_{i+1,j}$ is a linear function with regard to $X_i$, we can derive $\mu_j$ and $\sigma_j^2$ using the linear regression method with least square (LS), which can be derived as
\begin{small}
\begin{equation}
\displaystyle{\sigma_{j}^{2}=\frac{2\left[\sum\limits_{i=0}^{M-1} X_{i}^{2}-\left(\sum\limits_{i=0}^{M-1} X_{i}\right)^{2}\right]}{\left[\sum\limits_{i=0}^{M-1} X_{i}\left(D_{i, j}-D_{i+1, j}\right)\right]\left[\sum\limits_{i=0}^{M-1} X_{i} \sum\limits_{i=0}^{M-1}\left(D_{i, j}-D_{i+1, j}\right)\right]}.}
\end{equation}
\end{small}

Considering the symmetry of constellation points, we have $E(X_i )=0$, and $\sigma_j^2$ can be written as
\begin{equation}
\displaystyle{\sigma_{j}^{2}=\frac{2 \sum_{i=0}^{M-1} X_{i}^{2}}{\sum_{i=0}^{M-1} X_{i}\left(D_{i, j}-D_{i+1, j}\right)}.}
\end{equation}
Finally $\mu_j$ can be solved and given by
\begin{equation}
\displaystyle{\mu_{j}=\sum_{i=0}^{M-1} X_{i}-\frac{\sigma_{j}^{2}}{2} \sum_{i=0}^{M-1}\left(D_{i, j}-D_{i+1, j}\right)+1.}
\end{equation}
It should be noted that the distribution of $P(s_j |\mathbf{y})$ is fitted through the sampled paths instead of statistical calculation $\mu_i=\sum_{s_i} p(s_i |\mathbf{y}) s_i$ and $\sigma_i^2=\sum_{s_i} p(s_i |\mathbf{y}) |s_i-\mu_i|^2$.  The reason is that Gaussian distribution itself is a continuous distribution. After discretizing the Gaussian distribution, the mean and variance obtained from discrete sample points are not consistent with the parameters in the original Gaussian distribution. When the sampling is sparse, the difference between them is even great. Because the number of demodulation path samples is relatively small, the distribution is relatively concentrated, especially when SNR is high. In this way, the statistical variance is much smaller than $\sigma^2$, which would lead to calculation error in LLRs. Therefore fitting the distribution of $p(s_j |\mathbf{y})$ is necessary in these scenarios.

\subsection{Approximate Gaussian distribution transformation}

The assumption that the marginal distribution $P(s_j |\mathbf{y})$ conforms to Gaussian distribution is based on continuous integration of $P(\mathbf{s} |\mathbf{y})$. Strictly speaking, for discrete summation problem in (6), the distribution of $P(s_j |\mathbf{y})$ is not exact Gaussian distribution. Also, in practical tree searching algorithm, the sampled metrics derived from $P(s_j |\mathbf{y})$ also deviate from Gaussian distribution due to limited paths. When the number of layers is high and the correlation between layers is strong, assumption may not hold where multiple peaks will appear. In this subsection we propose a method to transform the actual sampling distribution, making it meet the Gaussian distribution, and then extract the moments on the transformed distribution.

In particular, we define the following mathematical problem:

\textbf{\textsl{Distribution Transformation Problem:}} Let the discrete a posteriori probability distribution vector obtained by sampling be $\mathbf{p}$, we need find a transport transformation on distribution $T(\mathbf{p})$ with one-to-one mapping, so as to minimize the KL divergence of $T(\mathbf{p})$ and $\mathbf{q}$, where $\mathbf{q}$ is the discrete samples of Gaussian distribution. Namely
\begin{equation}
\displaystyle{\min _{T} D_{KL}(\mathbf{q} \| T(\mathbf{p})).}
\end{equation}

The problem in (15) is a Monge problem of optimal transport theory [19]. Considering that the complexity of the transformation should be as low as possible, here we restrict the transformation on the distribution $\mathbf{p}$ to the transformation on the constellation point $\mathbf{s}$ so as to minimize the KL divergence of $\mathbf{p}$ and $\mathbf{q}$. Here we present the following theorem:

\textbf{\textsl{Theorem 1:}} Suppose the discrete a posteriori probability distribution vector obtained by sampling is $\mathbf{p}=[p_1,p_2,\cdots,p_K]$, where $K$ is the number of samples, then the transformation on distribution $T(\mathbf{p})=[p_{(1)},p_{(2)},\cdots,p_{(K)}]$ is the optimal transport transformation of the original Monge problem that satisfies (15). Here $p_{(i)}$ denotes the order statistics of $p_i$.

\textbf{\textsl{Proof:}} According to the definition of KL divergence
\begin{equation}
\displaystyle{\min _{T} D_{KL}(\mathbf{q} \| T(\mathbf{p}))=\min _{T}\left(\sum_{i=1}^{2^{M}} q_{i} \log \frac{q_{i}}{T\left(p_{i}\right)}\right),}
\end{equation}
where $\mathbf{q}=[q_1,q_2,\cdots,q_K]$, we can get
\begin{equation}
\displaystyle{\min _{T} D_{KL}(\mathbf{q} \| T(\mathbf{p}))=\min _{\mathbf{T}}\left(\sum_{i=1}^{2^{M}} q_{i} \log q_{i}-\sum_{i=1}^{2^{M}} q_{i} \log T\left(p_{i}\right)\right).}
\end{equation}
Here the first term of (17) is independent with $T(p_i)$, so the minimization of $D_{KL}(\mathbf{q} \| T(\mathbf{p}))$ is equivalent with maximization of $\sum_{i=1}^{2^{M}} p_{i} \log T\left(p_{i}\right)$. In the light of sequence inequality, we can prove that
\begin{equation}
\displaystyle{\sum_{i=1}^{2^{M}} q_{i} \log T\left(p_{i}\right) \leq \sum_{i=1}^{2^{M}} q_{(i)} \log p_{(i)},}
\end{equation}
where $p_{(i)}$ denotes the order statistics of $p_i$ [20]. (18) means the optimal transport transformation is to rearrange the original distribution into the same order of a desired Gaussian distribution. $\hfill\blacksquare$

Considering that the larger probability value plays a leading role in the calculation of mean and variance, a part of the smaller probability value can be omitted when fitting the mean and variance. It can reduce the computational complexity in two aspects: on the one hand, it can reduce the complexity of computing the mean and variance; on the other hand which is more importantly, it can reduce the requirement for the number of sampling paths. We have the following theorem:

\textbf{\textsl{Theorem 2:}} For the sampled marginal posterior probability subject to any multimodal distribution, the minimum number of sampling paths to derive the means and variances for all layers is $4N_t+1$ for $N_t$-layer data streams.

\textbf{\textsl{Proof:}} By denoting $f(X_i)=\min _{\mathbf{s} \in\left\{s_{j}=X_{i}\right\}}\|\mathbf{y}-\mathbf{H} \mathbf{s}\|^{2}$, (9) can be rewritten as
\begin{equation}
f(X_i)=\frac{\left|X_{i}-\mu_{j}\right|^{2}}{2 \sigma_{j}^{2}} +\frac{1}{2} \log \left(2 \pi \sigma_{j}^{2}\right)+\min _{\mathbf{s} \in \mathcal{S}}\|\mathbf{y}-\mathbf{H s}\|^{2}.
\end{equation}
Suppose we have collected the shortest paths under three adjacent constellation points in a certain layer, which are recorded as $f(X_{i-1})$, $f(X_{i})$ and $f(X_{i+1})$. Then the differential equation of (19) is given by
\begin{small}
\begin{equation}
\begin{aligned}
{f\left( X_{i} \right) - f\left( X_{i-1} \right) = \frac{1}{2\sigma_j^{2}}\left( {X_{i}^{2} - X_{i-1}^{2}} \right) + \frac{1}{\sigma_j^{2}}\left( {X_{i-1} - X_{i}} \right)\mu_j}, \\
{f\left( X_{i} \right) - f\left( X_{i+1} \right) = \frac{1}{2\sigma_j^{2}}\left( {X_{i}^{2} - X_{i+1}^{2}} \right) + \frac{1}{\sigma_j^{2}}\left( {X_{i+1} - X_{i}} \right)\mu_j}. \\
\end{aligned}
\end{equation}
\end{small}
Equation (20) only has two unknowns. It is a system of binary linear equations with one order, and there is a unique solution for (20). For the case of $N_t$ layers, the global minimum path is the same for all layers. For each layer, the required sampling paths are the minimum paths within constellation points adjacent to the minimum points in real and imaginary parts, which are four nodes at each layer. Hence at least $4N_t+1$ paths are required to compute the mean and variance. $\hfill\blacksquare$

\subsection{Learning based LLR recovery design}

\begin{figure} [t]
\centering
\includegraphics[width=0.5\textwidth]{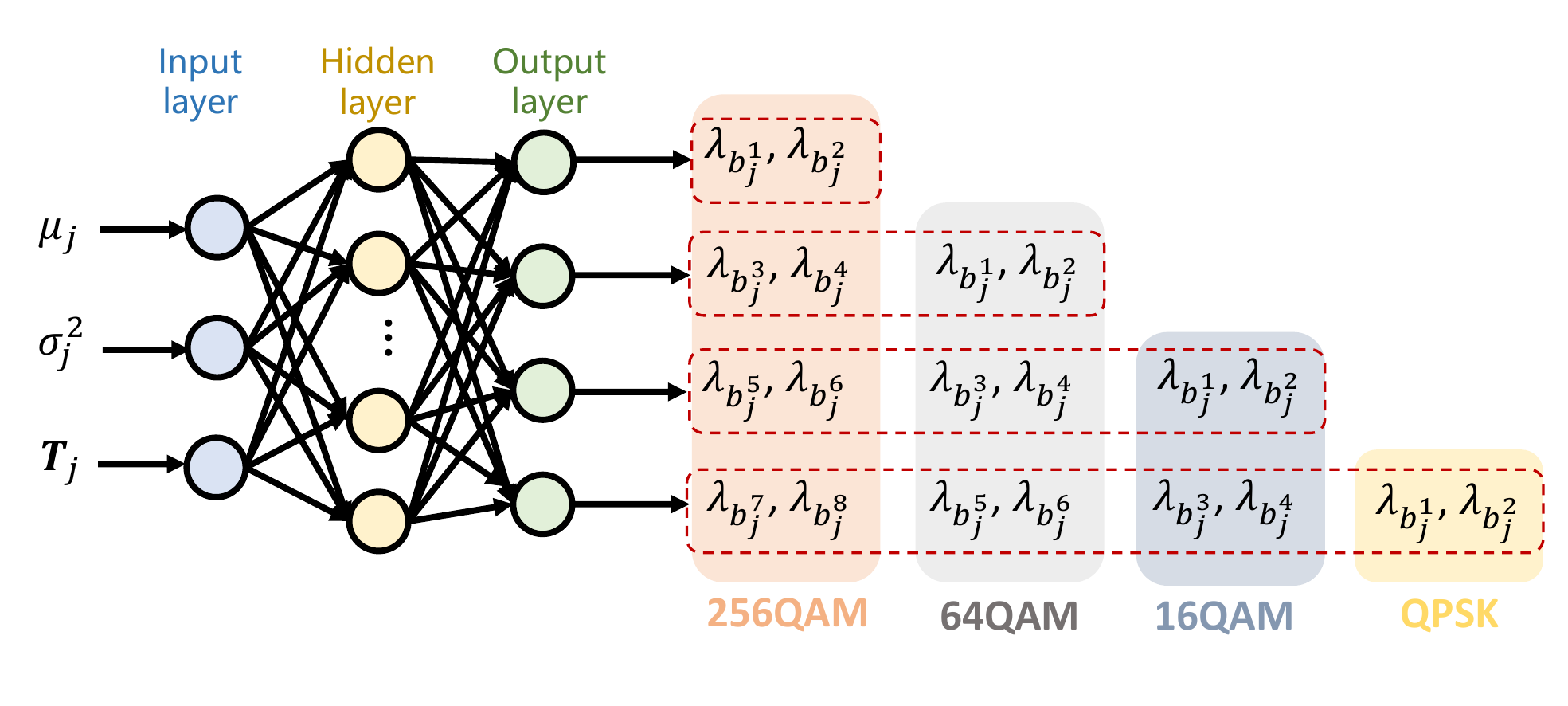}
\caption{Low-complexity LLR recovery for M-QAM.}
\label{fig}
\end{figure}

In this section, we propose a machine learning module to recover the LLR from the derived mean and variance. This module can establish the relationship between extracted moments (mean and variance) and LLR, without the need to recover ${p}\left(s_{j} \mid \mathbf{y}\right)$ explicitly. The lightweight neural network has two advantages: (i) it acts as a classifier to automatically recognize the scenarios where LLRs need to be modified (such as high correlation between layers or missing counter-hypotheses); and (ii) it acts as a regression filter to automatically learn the relationship between the extracted moments (mean and variance) and optimal LLRs, which acts as labels derived from the practical data with full-sampling. Moreover, MPPS can reduce the complexity of log-MAP calculations. The network architecture is shown in Fig. 1.

The lightweight network composes of a lightweight network with one hidden layer. The network input is the received first- and second-order moments $\{\mu_j, \sigma_j^2\}$ and the optimal transformation $T_j$, and the output is the LLRs of all bits in a symbol. One of the advantages of the neural network compared with the model-based algorithm is that it can learn the practical scenarios with measured data, and to identify the scenarios whether the LLR needs to be modified.

The network implementation can be divided into two steps. The first step is to train the constructed fully-connected network, adjust the weight of the network, and mine the nonlinear relationship between the input vector and the ideal output vector. The second step is to use the trained network to demodulate the signals. In our experiment settings, we choose the $L_2$ loss as the cost function.

Now we can summarize the MPPS algorithm into the following steps:

\textbf{Step 1:} For each layer $j$, calculate $D_{i, j}=\min _{\mathbf{s} \in\left\{s_{j}=X_{i}\right\}}\|\mathbf{y}-\mathbf{H s}\|^{2}$ of the sampled paths, sort $D_{i, j}$s into a specific order where their marginal posterior probabilities satisfy the optimal transport transformation $T(\mathbf{p})$ by Theorem 1.

\textbf{Step 2:} Calculate $\mu_j$ and $\sigma_j^2$ of the marginal posterior probability according to (16) and (17).

\textbf{Step 3:} $\mu_j$ and $\sigma_j^2$ are fed into a lightweight neural network to fit the log-MAP LLR with
$\lambda_{b_{j}^{i}} = g(\mu_j, \sigma_j^2, T_j)$, where $g$ is realized by the network.

\section{Simulation Results}


\begin{table}[t]
\centering
\caption{Key Parameters in Simulations}
\label{table}
\begin{threeparttable}
\begin{tabular}{c||c}
\hline
\textbf{Parameter} & \textbf{Value} \\
\hline
Carrier frequency & 2.15 GHz \\
Subcarrier spacing & 15 kHz \\
Modulation & 64 QAM \\
Num. of antennas & Tx 4, Rx 4 \\
Num. of resource blocks & 52 \\
Num. of layers & 4 \\
Channel & AWGN, TDL-A \\
Decoder & low-density parity-check (LDPC) \\
Code rate & 0.466 \tnote{1} \\
\hline
\end{tabular}
\begin{tablenotes}
\footnotesize
\item[1] The code rate is chosen according to 3GPP protocol [22]. Due to limited space, only one code rate is presented here. However, all the code rates have similar conclusions.
\end{tablenotes}
\end{threeparttable}
\end{table}

\begin{figure} [t]
\centering
\includegraphics[width=0.45\textwidth]{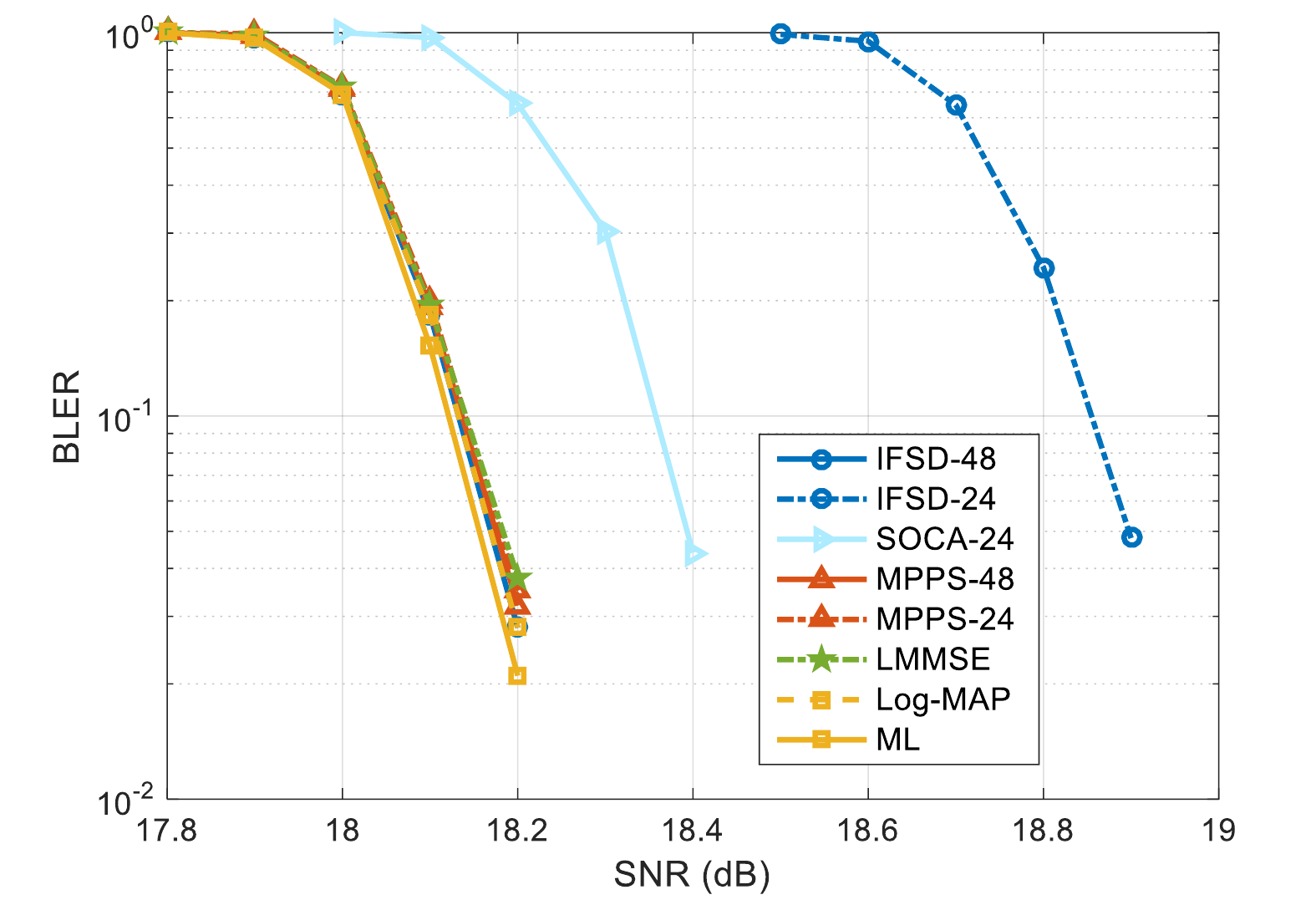}
\caption{BLER versus SNR for different algorithms with AWGN MIMO channel.}
\label{fig}
\end{figure}

We evaluated the performance of the proposed MPPS algorithm presented in the previous section in a 5G-NR simulation link. The compared algorithms are well-known schemes within each class of algorithms, such as linear detector (LMMSE), tree searching method (IFSD [3]), candidate adding algorithm (SOCA [9]), log-MAP detector [21], and ML detector. Different numbers of sampled paths for IFSD are tested with 4-layer data streams, Within the same path number, IFSD, SOCA and MPPS have the same set of sample paths in order to be fair. In particular, the number of sampling path of log-MAP is set to 48 corresponding to the largest path number of IFSD in our simulation. The BLER is used to measure the performance of these detectors. Other simulation parameters can be seen in Tab. I.


First, we have tested the performance of different algorithms in a channel with additive white Gaussian noise (AWGN). The SNR of the system, defined as
\begin{equation}
\displaystyle{SNR=\frac{\mathbb{E}\|\mathbf{H s}\|^{2}}{\mathbb{E}\|\mathbf{n}\|^{2}},}
\end{equation}
is used to measure the noise level. From Fig. 2 we can see that the proposed MPPS has comparable performance as ML, and outperforms the IFSD algorithm. Specifically, for BLER=0.1, MPPS-24 achieves 0.8 dB and 0.3 dB SNR gain compared with IFSD-24 and SOCA-24 algorithm, respectively. Note that for 24 paths, IFSD has performance degradation because not enough paths are selected for counter-hypotheses paths, and the performance gain is achieved for MPPS due to immunization of missing counter-hypotheses.

\begin{figure} [t]
\centering
\includegraphics[width=0.45\textwidth]{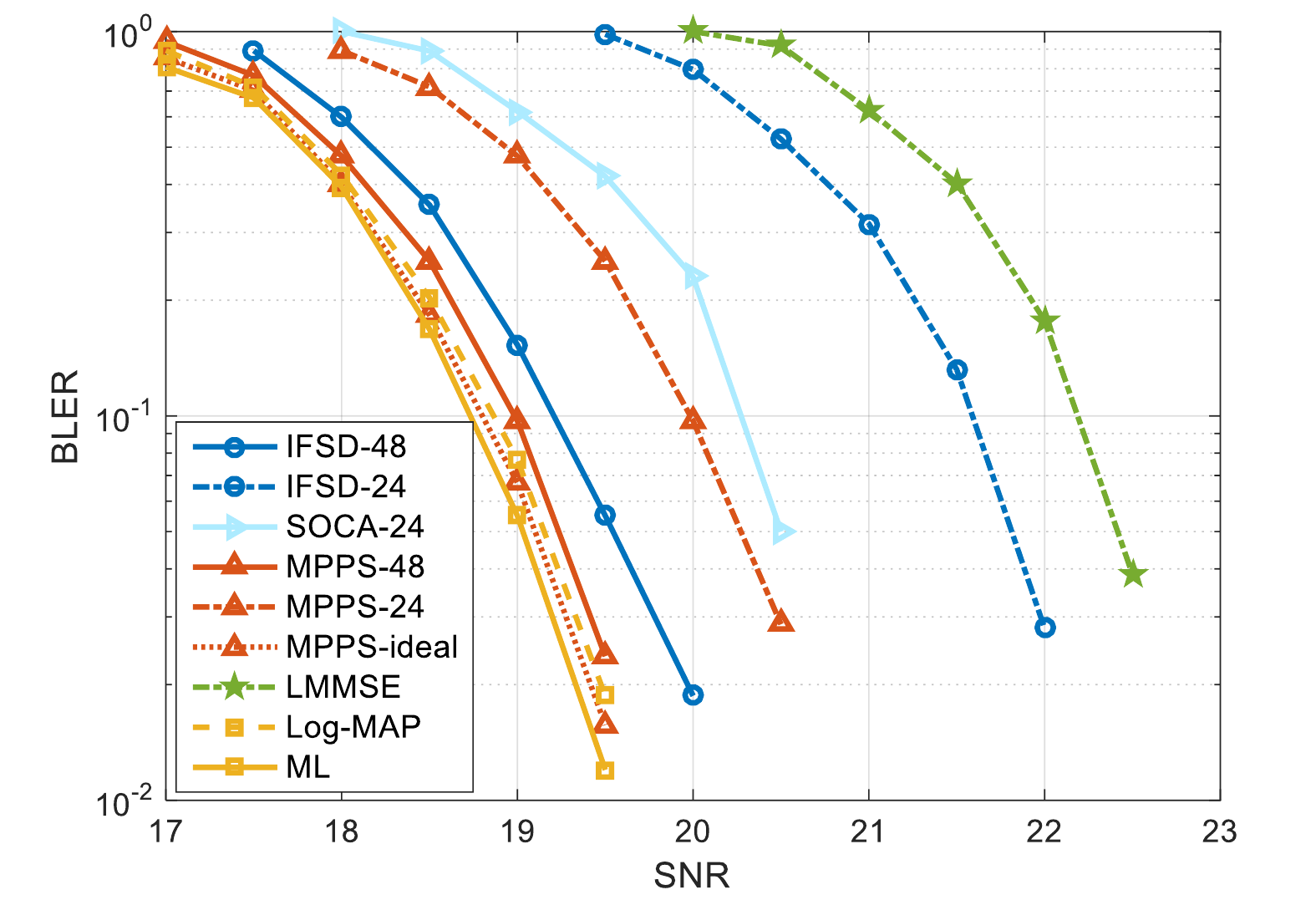}
\caption{BLER versus SNR for different algorithms with fading MIMO channel.}
\label{fig}
\end{figure}
\begin{figure} [t]
\centering
\includegraphics[width=0.45\textwidth]{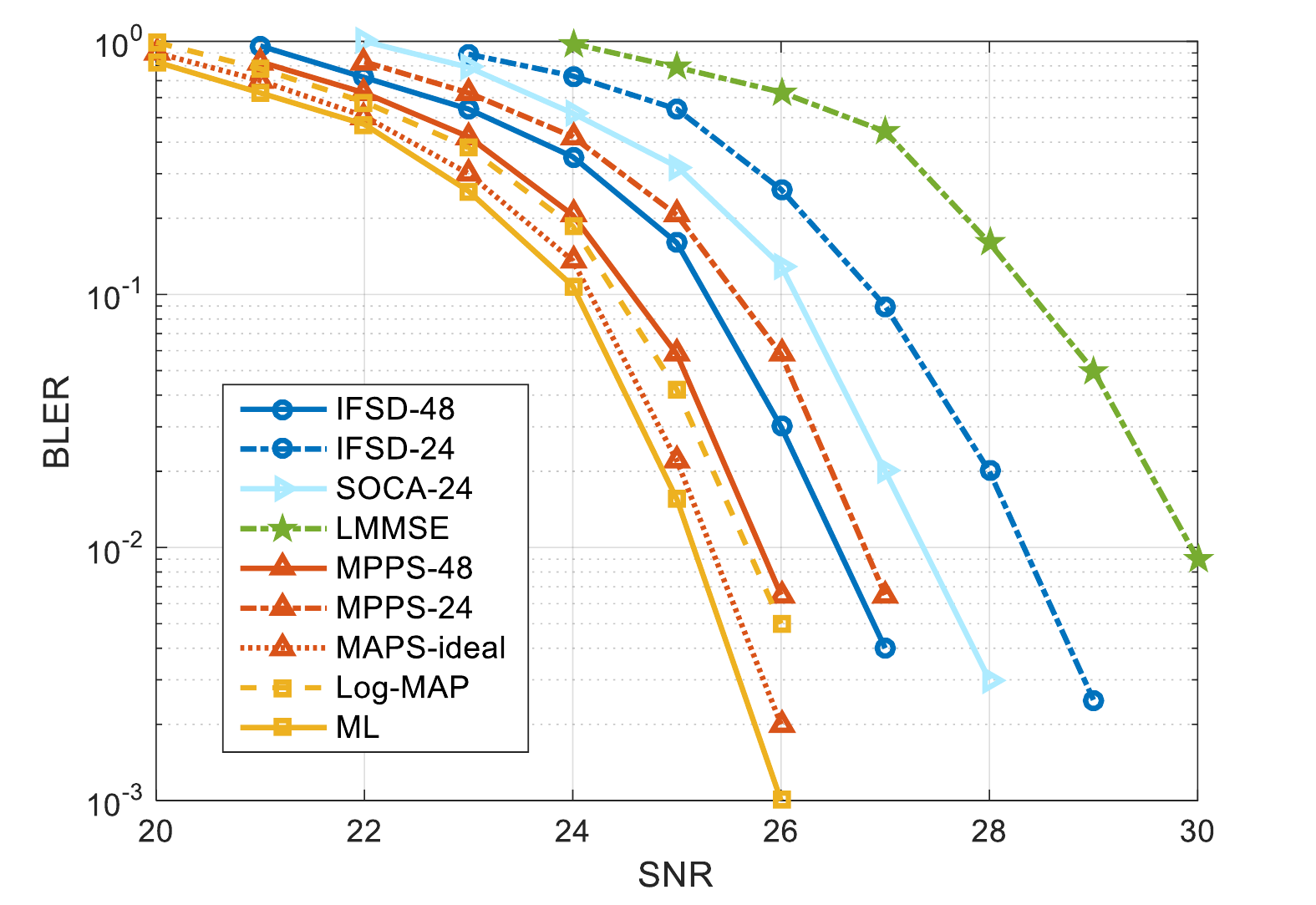}
\caption{BLER versus SNR for different algorithms with fading correlated MIMO channel.}
\label{fig}
\end{figure}


Fig. 3 compares the average BLER performance of different algorithms under the fading MIMO channel. Here the channel is adopted as time delay line (TDL) channel defined in [22], and performance of TDL-A is shown as a representation with a mobile station moving at 30 km/h speed. It can be seen that MPPS-48 can outperform the other the other competition algorithms and approach the log-MAP and ML. For BLER=0.1, MPPS-48 attains SNR gain of 0.3dB and 3.2dB for IFSD-48 and LMMSE respectively. Note that according to Theorem 2, MPPS with 17 sampled paths can approach ML performance due to exact moments calculation of marginal posteriori probability (see the curves named MPPS-ideal). In practice, we need a little more paths to increase the probability that the shortest path is selected. Although the performance of MPPS-24 is a little worse than MPPS-48 due to the decreased probability to find the shortest path, it still attains SNR gain of 1.6dB and 2.2dB for IFSD-24 and LMMSE respectively. The reason of such gain comes from two aspects. In addition to being robust to missing counter-hypothesis issue, the neural network in MPPS fits the log-MAP calculation.


Finally we consider the correlated MIMO channel in this section, which can be described by the Kronecker model
\begin{equation}
\displaystyle{\mathbf{H}=\mathbf{R}_R^{1/2} \mathbf{A} \mathbf{R}_T^{1/2},}
\end{equation}
where $\mathbf{A}$ is the fading channel matrix, $\mathbf{R}_R$ and $\mathbf{R}_T$ are the spatial correlation matrix at the receiver and transmitter respectively, which are generated according to the exponential correlation model [23] where correlation coefficients for both transmitting and receiving antennas adopt 0.3. Here the channel model adopts TDL-A model as well. Fig. 4 illustrates the BLER performance of different algorithms under correlated TDL-A MIMO channels. In that case, the MPPS can obtain more performance gain even if the channel has correlation. Compared with 0.3dB and 1.6dB gain obtained under the independent fading channel, MPPS can obtain 1.5dB and 2.2dB SNR gain under the correlated MIMO channel. It should be noted that although ML can achieve 0.5dB gain over MPPS-48 under correlated channel, the complexity of ML is unacceptable as shown in Fig. 5. Hence MPPS achieves near-optimal performance under limited complexity.


We have also compared the complexity of different algorithms based on the CPU time, as shown in Fig. 5. All the algorithms are implemented on a standard 3.00GHz Intel Xeon E5-2690 CPU using C++. It can be seen that MPPS has lower complexity compared with other algorithms with the same sampled paths. In particular, the complexity of MPPS-24 is lower than all the other algorithms except LMMSE. Since MPPS is realized through a simple moments calculation approach with reduced number of required samples, it has fewer floating-point operations than conventional tree search algorithms and therefore yields a lower computational complexity.

\begin{figure} [t]
\centering
\includegraphics[width=0.5\textwidth]{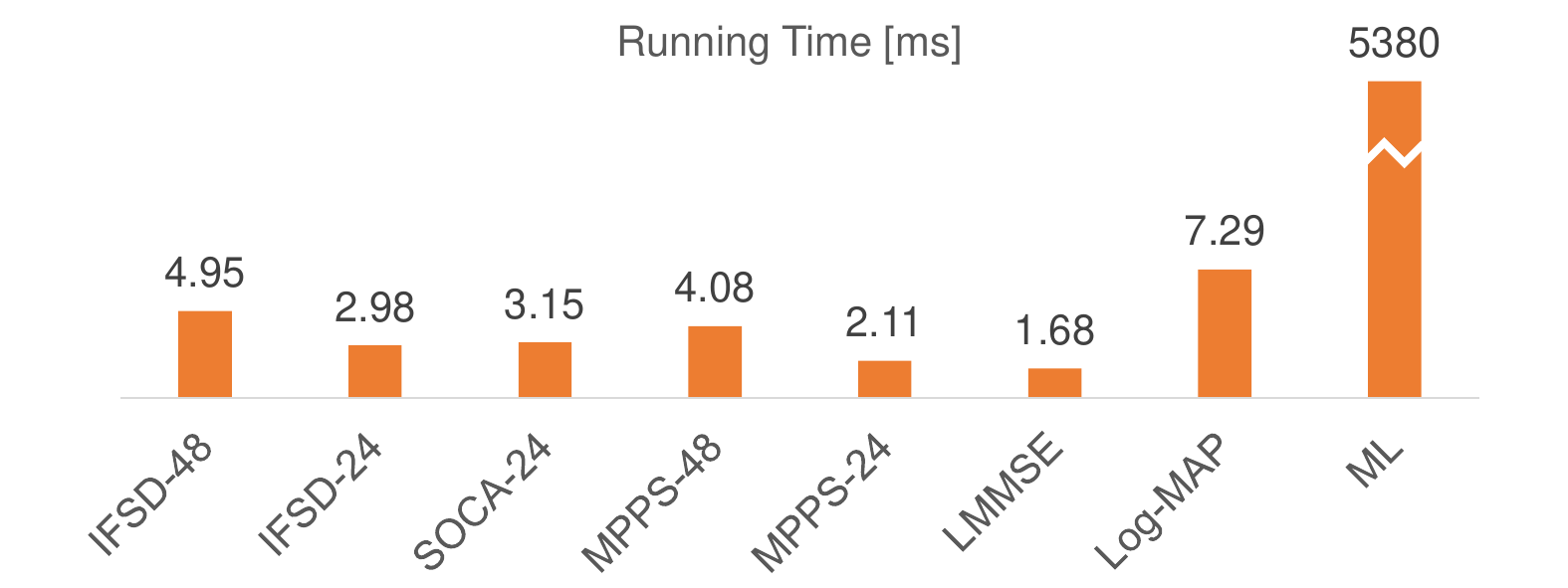}
\caption{Computational complexity comparison of the demodulation schemes per received symbol.}
\label{fig}
\end{figure}

\section{Conclusion}

This paper proposes a novel soft MIMO detection algorithm under the criterion of marginal posteriori probability estimation. Unlike traditional algorithms based on the quasi-ML search, we fit the marginal posteriori probability distribution by a small number of sampled paths, and estimate the statistical information of the marginal distribution, with reduced samples and free from missing counter-hypotheses. We also propose a lightweight learning approach to recover the log-MAP LLR using the estimated statistical information with low complexity. Simulation results show that the proposed algorithm can achieve significant performance gain compared with traditional detectors with a relatively low computational time.

\end{document}